\documentstyle[preprint,aps]{revtex}

\draft  \title{Magneto-polarisability of mesoscopic rings }

\author{Y. Noat, B. Reulet, and H. Bouchiat}

\address{Laboratoire de Physique des Solides, Associ\'e au CNRS, B\^at 510,
Universit\'e Paris--Sud, 91405, Orsay, France.}

\begin{document}

\maketitle

 \begin{abstract} We calculate the average polarisability of two dimensional
mesoscopic rings in the presence of an Aharonov-Bohm flux. The screening is
taken into account self-consistently  within a mean-field approximation. We
investigate the effects of statistical ensemble, finite frequency and disorder.
We emphasize geometrical effects which make the observation of field dependent
polarisability much more favourable on rings than on disks or spheres of
comparable radius. The ratio of the flux dependent to the flux independent part
is estimated for  typical GaAs rings.

 \end{abstract}

Transport properties of mesoscopic metallic conductors are known for a long
time to be quite sensitive to the quantum phase coherence of  electronic wave
functions at low temperature. More recently it has been shown both
theoretically and experimentally \cite{houches}, that the orbital magnetism
also presents remarkable signatures of this phase coherence. The purpose of the
present paper is to discuss  to what extend the electrical polarisability i.e.
the response of a metallic  sample to an  electrostatic field is also sensitive
to mesoscopic effects. This quantity can be  experimentally measured by
inserting the mesoscopic samples   into a capacitor. In the linear response
regime, the change in the capacitance is directly proportional to the average
polarisability $\alpha $ of the particles. If the typical size $L$ of the
particle is much larger than the Thomas Fermi screening length $\lambda_s$,  
the polarisability is mostly determined by the geometrical shape of the
particle with a small correction of order  $\frac{\lambda_s}{L}$ \cite{rice73}.
 If furthermore  $L\preceq L_\phi $ the phase coherence length,  one can
expect to find a  quantum correction $\delta \alpha _Q $ to the polarisability:
$ \alpha = \alpha _{TF}+\delta \alpha _Q $. The {\it typical value}  of this
magnetic field dependent quantum correction was estimated from diagrammatic
theory  \cite{berko92} in disks and spheres.  We have calculated the {\it
average} of the magneto-polarisability  of {\em \ two dimensional mesoscopic
rings} in the presence of an {\em Aharonov-Bohm flux}, with in-plane electric
field. In this geometry $\alpha _{TF}$ is of the order of $R ^3$ and the
quantum correction is of the order of $\lambda _s/W$, where $R$ and $W$ are 
the radius and width of the ring. This correction is much larger than for disks or
spheres of comparable radius with approximatively the same value for $ \alpha =
\alpha _{TF}$.  We consider a time dependent electric field which leads to two
different limits whether the frequency is smaller or larger than the inverse
inelastic time $\gamma $. Nevertheless, both $\omega $ and $\gamma $ are 
supposed to be small compared to the mean level spacing  $\Delta $.  We will 
consider the average value of the polarisability in grand canonical ensemble
(GCE) as well as in the canonical ensemble (CE). In this letter we present
mainly analytical results which have been confirmed by numerical simulations,
to be published separately \cite{noat96}. The agreement between our findings and
the recent results of Efetov on disks and spheres \cite{efe96} is only
qualitative, (in  particular, we strongly disagree on the disorder dependence
of the effect.)

We have treated the screening of the    applied electric field  by the
electrons within a mean field approximation which leads to the one electron
Hamiltonian in an Aharonov-Bohm ring ($\vec A$ is the static vector potential):

\begin{equation}   H(t)=\frac{(-i\overrightarrow{\nabla }+e\overrightarrow{A}
)^2}{2m}+V_{dis}(\overrightarrow{r})+e \Phi
_{scr}(\overrightarrow{r})\exp(i\omega t) +V_{int}(\overrightarrow{r})  
\end{equation}

Where $\Phi _{scr}(\overrightarrow{r})= E_0F (\overrightarrow{r} )$  is the 
effective screened    potential experienced by the electrons in the time
dependent external electric field $\vec E_0\exp(i\omega t)$  which is assumed
to be small enough  in order that linear response is valid .  $V_{dis}$ is a
disordered impurity potential and $V_{int}$ is the mean field  potential due to
electron-electron interactions.  In an Aharonov-Bohm geometry  $\Phi_{scr}
(\overrightarrow{r})$  and $V_{int}(\overrightarrow{r})$ depend on the magnetic
flux, as pointed out by Bttiker  \cite{butti94}. In the present work we
neglect  $V_{int}(\overrightarrow{r})$ and consider consider  single electron
eigenstates and eigenvalues, but we fully  take into account of the screening.
$\Phi_{scr} (\overrightarrow{r})$ is  related  self consistently to  the field
induced  shift of the electronic density $\delta n (\overrightarrow{r},E_0)$,
according to:

\begin{equation} \left\{  \begin{array}{lcl}  \Phi_{scr}
(\overrightarrow{r})=\displaystyle\int\frac{e\delta n(\vec{r'})} { 4\pi
\epsilon_0|\vec{r'}-\vec{r}|}d\overrightarrow{r'}
+\overrightarrow{E_0}.\overrightarrow{r}  & \\ \delta
n(\overrightarrow{r})=\int \chi(\overrightarrow{r},\overrightarrow{r'},\omega)
\Phi_{scr}  (\overrightarrow{r'})d\overrightarrow{r'} \end{array} \right
.\label{eqTF} \end{equation}

 The ac electric field induces a dipolar moment which is given by: $  \delta
P_x(t)\equiv e Tr\left[ (\rho (t)-\rho _0)X\right] \equiv \alpha E  $. Where Tr
denotes the trace over all states. $X$ is the projection of the position
operator along the direction of the electric field and  $\rho $ is the density
matrix, solution of the master equation  in the presence of the time dependent
perturbation:

\begin{equation}  \displaystyle i\frac{\partial \rho }{\partial t}=\left[
H,\rho \right] -i\gamma (\rho-\rho _{equi})  
	\nonumber
	\end{equation}
	  $\gamma$ is the inverse relaxation time towards equilibrium. $\rho _{eq}$
is the instantaneous equilibrium density matrix which satisfies: $ \left[
H,\rho _{eq}\right] =0  $. Following the same steps than for the computation of
the orbital susceptibility of Aharonov Bohm rings,  within linear  response
theory \cite{trive88,reule94}, the polarisability can be expressed as a
function of eigenvalues $\epsilon _\alpha$ of the unperturbed Hamiltonian, 
matrix elements of the operators $X$ and $F(\vec r)$ taken between the
corresponding eigenstates, and populations $f_\alpha =\displaystyle \left(\exp
\left[ \frac{ (\epsilon _\alpha -\mu )}{k_B T} \right] +1\right )^{-1}$

 \begin{equation} 
	\alpha =2e^2\left\{\sum_ {\alpha\neq\beta} \displaystyle \frac{f\alpha
-f_\beta }{\epsilon _\alpha -\epsilon _\beta }\frac{\epsilon _\alpha -\epsilon
_\beta -i\gamma }{ \epsilon _\alpha -\epsilon _\beta +\omega -i\gamma }
X_{\alpha, \beta}  F (\overrightarrow{r},\phi )_{ \beta,\alpha}+ \frac
\gamma {\gamma +i\omega }\sum_\alpha \frac{\partial f_\alpha }{\partial
\epsilon _\alpha } X_{\alpha, \alpha}  F (\overrightarrow{r},\phi )_{\alpha,
\alpha} \right\}   \label{pola-general} 
	\end{equation}

 In the case where the dimensions of the sample are at least a few times larger
the  screening length,  $\lambda_s \simeq \lambda _F$ it is reasonable to
describe the screening by a  Thomas-Fermi approximation,   \cite{rice73}  where
the  induced electronic density is proportional to the  potential. For a 2D 
system: $\chi_{TF}(\overrightarrow{r},\overrightarrow{r'})= (2\pi\epsilon_0
/\lambda_s) \delta( \overrightarrow{r}-\overrightarrow{r'})$. The  potential in
the ring  has then the form \cite{noat96}:  $ F^{TF} (\overrightarrow{r})=
f(r,\lambda _s,W)\cos  \theta  $, where the function $f(r)$  depends on the
aspect ratio of the ring. For $W<<R$   the charge density is nearly equally
distributed between the 2 boundaries of the ring, on a length of the order
$\lambda _s$, otherwise the charge is concentrated on the external boundary. 
The Thomas-Fermi polarisability is related to $f(r)$ by: $ \alpha _{TF}
=\displaystyle\frac{2\epsilon_0\pi^2}{\lambda_s}\int r^2 f(r) dr  $. For small
$\lambda _s$, we recover the polarisability obtained by solving the Poisson
equation, whose value depending on the ratio $W/R$ extrapolates between the
value of the 1D ring $\alpha_{1D}= \displaystyle \frac{\epsilon_0
\pi^2R^3}{ln(R/W)}$ and the value of the disk \cite{landau}
$\alpha_{2D}=\displaystyle\epsilon_0 \frac {16}{3} R^3$.  As, the ratio
$\lambda_s/W$ increases, size dependent corrections decrease the value of $
\alpha _{TF}$ .

	In order to estimate the quantum correction to the polarisability let us
write the local electric susceptibility tensor as following: $\chi
(r_1,r_2)=\chi_{TF} (r_1,r_2)+\delta \chi(r_1,r_{2 },\phi)  $.  Where  $\delta
\chi $ is the flux dependent quantum correction. The same for the average
screened potential: $ F (\overrightarrow{r},\phi )=F^{TF} (
\overrightarrow{r})+\delta F (\overrightarrow{r}, \phi ) $. In first order of
$\delta \chi$ and $\delta F$ which are self consistently related through
eq.\ref{eqTF}  the quantum correction to the polarisability in the limit $(\omega,\gamma << \Delta)$ can then be
expressed as:

\begin{equation} \delta \alpha (\phi) =2e^2 \delta  \left ( \sum_{\alpha \neq \beta}
 \frac{f_\alpha -f_\beta }{\epsilon _\alpha -\epsilon _\beta }| F_{\alpha
,\beta}^{TF} (\overrightarrow{r })| ^2+ \frac {\gamma} {\gamma +i\omega } \sum
_\alpha \frac{\partial f_\alpha }{\partial \epsilon _\alpha }  |F_{\alpha
,\alpha}^{TF} (\overrightarrow{r})| ^2 \right) \label{alpha-gene}
\end{equation}

 Note that  this expression is very similar to the unscreened polarisability
\cite{gorko65,frahm90,entin90}  where the matrix elements of the operator $X$
are replaced by those of the average, flux independent, screened potential.
  For the discussion of this quantity in the zero frequency limit it will be
important to distinguish between CE and GCE averages as well as between the
order of the limits $\gamma >>\omega \rightarrow 0$ (thermodynamic
polarisability $\alpha _T$) and $\omega >> \gamma \rightarrow 0$ (dynamical
polarisability $\alpha_D$). On this point we disagree with the statement of
Efetov who claims that all the limits should be identical\cite{efe96}. In order
to obtain the following  results valid at zero temperature, we have used the
expressions of the CE and GCE average population factors $<f_\alpha- f_\beta>$
and $<\partial f_{\alpha} / \partial \epsilon _{\alpha}>$ which has been
extensively studied in \cite{reule94,kamen94,kamen95} in the
context of the ac magnetic response of Aharonov-Bohm rings. We call $\delta
\alpha$ the flux dependent part of $\alpha$.

 In the GCE:

\begin{equation} \left\{  \begin{array}{lcl}  \delta \alpha_T^{GCE} =\delta
\left(\frac {2^2}{E_F} Tr(F^{TF})^2\right)=0   & \\ \delta \alpha_D^{GCE}
=\frac {2e^2}{E_F} \delta \left(\sum_{\alpha \neq \beta
}|F_{\alpha,\beta}^{TF}|^2 \right)= -\frac {2e^2}{E_F}\delta \left(
\sum_{\alpha}|F_{\alpha,\alpha}^{TF}|^2 \right) \end{array} \right.
\label{polGCED}  \end{equation}

 Where $\Delta$ is the average level spacing and $E_F$ is the Fermi energy.  
In the CE at zero temperature, the quantities $\partial f_\alpha / \partial
\epsilon_\alpha=0 $, there is no contribution of diagonal matrix elements to
the polarisability. The thermodynamic and dynamical polarisability  are
identical.

 \begin{equation}  \delta \alpha_T^{CE}=\delta \alpha_D^{CE}=\frac
{e^2}{E_F}\delta  \left (\sum_{\alpha \neq \beta}|F_{\alpha,\beta }^{TF}|^2 (
1+\frac{\Delta}{\epsilon _\alpha -\epsilon _\beta})\right)  \label{polCE} 
\end{equation}

    Expression (\ref{polGCED}) shows that the flux dependent polarisability in
the GCE is directly related to the matrix elements of the screened potential
$F^{TF}$ which  can be estimated from the Fourier spatial decomposition:  $
\displaystyle F^{TF}(\vec r ) = \sum_ {n=-M/2} ^{M/2} F_n \exp(i\vec q_n \vec
r)$   where $\vec{q_n} =\displaystyle \frac  {n\pi}{W} \vec{u_r}+\displaystyle
\frac {1}{R}\vec u_{\theta} $ and $M$ is the number of transverse channels in
the ring.  Using the  general relation established by McMillan \cite{macmi} for
matrix elements of $\exp(i\vec q\vec r)$:

\begin{equation} |\exp (i\vec{q} \vec r)_{\alpha,\beta }|^2= \frac{\Delta}{\pi}
\frac{D q^2}{ D^2q^4+(\epsilon_\alpha-\epsilon _\beta)^2} 
	\label{eq:c}
	\end{equation} 
	 $D$ is the diffusion coefficient, for low energy difference the main
contribution is given by the smallest
	$\vec q$ vector $\vec q_0$ for which $F_0=(\int f(r)dr)/W= 8 R \lambda_s
/3\pi^2W \sim 2 R/M $ (taking the 2D limit for $f(r)$). As a result:

\begin{equation}
	<|F_ {\alpha,\beta}^{TF}(\vec r)|^2> = \frac{\Delta}{\pi E_{c}}\left ( \frac
{8R\lambda_s}{3\pi^2 W} \right )^2 \label{eqPhie}
	\end{equation}$E_c=h/\tau_D=hD/(2 \pi R)^2$  is the Thouless energy and
$\tau_D $ is the diffusion time around the ring. This relation allows the
generalization of the properties of the $X$ operator established in
\cite{gorko65,sivim87}. More interesting for our purpose, these matrix
elements are also flux dependent. This flux dependence is  $\phi_0/2$ periodic
and presents a minimum at $\phi_0/4$, in contrast  with the average square of
the diagonal matrix element  which is maximum at $\phi_0/4$. The time reversal
symmetry property of the operator $F^{TF}(\vec r)$  implies that its diagonal
matrix elements  are real, even functions of flux  and can be developed in
successive powers of $\cos(2\pi \phi/\phi_0)$.  $|F_{\alpha,\alpha}^{TF}|^2 $
is maximum for multiple values of $\phi_0/2$  and minimum for $\phi_0/4
[\phi_0/2]$. The opposite flux dependence for the non diagonal matrix elements
directly follows from the flux independence of the trace  $ \sum_{\alpha ,
\beta}|F_{\alpha,\beta}^{TF}|^2=\sum{\vec r} F(\vec r)$. Note that flux dependences of the
diagonal and non diagonal matrix elements of the current operator which is an
odd function of flux, are just opposite as already noted in \cite{imry86}.  In
order to estimate quantitatively the flux dependence of
$|F_{\alpha,\alpha}^{TF}|^2$,  it is useful to  relate the diagonal matrix
element  $F_{\alpha,\alpha}^{TF}$ and the sensitivity of eigenvalues 
$\epsilon_\alpha$  to the screened  electric field: $
F_{\alpha,\alpha}^{TF}=\displaystyle \frac {\partial
\epsilon_{\alpha}}{\partial E_0}$.  Since electric field preserves time
reversal symmetry \cite{wilkin},  the typical derivative of the energy levels:
$<|\displaystyle \frac {\partial \epsilon_{\alpha}}{\partial E_0}|^2>$ is
proportional to $ 1/\beta$, where the parameter $\beta $ characterizes the
symmetry class of the Hamiltonian, determined by another parameter (flux, for
example), as a result: \begin{equation}  |F_{\alpha,\alpha}^{TF}|^2 (\phi_0/4)
(GUE ensemble)=\frac{1}{2} |F_{\alpha, \alpha}^{TF}|^2(0)(GOE ensemble) 
	\label{eqsym}
	\end{equation} 
	 From equations \ref{polGCED},\ref{eqPhie},\ref{eqsym} one deduces the
magneto-polarisability in the GCE:

\begin{equation}  \delta \alpha /\alpha= \frac{\alpha_D^{GCE}(\phi_0/4)-
\alpha_D^{GCE}(0)}{\alpha _0} =\frac8{3\pi^3}\frac{\Delta}{
E_{c}}\frac{\lambda_s}{W}=\frac8{3\pi^3}(\frac{\Delta \tau_D}{
h})\frac{\lambda_s}{W} \label{eq:}
	\end{equation} 
	  This result  implies a positive low field magneto-polarisability in
agreement with findings of Efetov \cite{efe96}.
  We also find an increase of the magneto-polarisability with disorder in the
diffusive regime,  in good agreement with numerical data depicted in figure
\ref{figxphi} but  at odd with reference\cite {efe96}, where a value of the
order of $(\frac{\Delta \tau_e}{ h})\frac{\lambda_s}{W}$ ($\tau_e << \tau_D$ is
the elastic scattering time) found for $\delta\alpha/\alpha$. A similar  $1/D$
dependence was also found for the typical value of the magneto-polarisability
\cite{berko92}.  It is also clear from our results that the ring geometry is
very favourable for the observation of quantum effects on the polarisability.
The extrapolation   for a disk or a sphere of same radius with comparable 
values of $\alpha_{TF}$ yields to $\delta \alpha / \alpha = \displaystyle
(\frac{\Delta}{ 2 E_{c}})\frac{\lambda_s}{R}$. The relative correction is thus
reduced  by a factor $(W/R)^2$ for the 2D disk  and a factor $ \lambda_F
W^2/R^3$ for the sphere. We can  estimate the relative magneto-polarisability
of a ring made from a GaAs/GaAlAs heterojunction with the following parameters:
$L=8\mu m$, $\lambda_s= 400$ \AA , $l_e= 2\mu _m$, $M=10$, $E_{c}=7\Delta$, to be
of the order of $3  10^{-3}$. Recent experiments measuring the ac complex
conductance of such Aharonov-Bohm rings coupled to an electromagnetic resonator
are sensitive enough to detect their magneto-polarisability \cite{rrbm95}. 

If we consider now the canonical ensemble (CE),  according to eq.\ref{polCE},
the flux dependent correction to the polarisability in the CE at $T=0$ depends
both on  matrix elements discussed above and energy levels. Assuming that
energy levels and matrix elements are indeed independent random variables, it
is possible to rewrite eq.\ref{polCE}  

\begin{equation} \delta \alpha^{CE}= \delta\left [ \alpha^{GCE} +\frac{
e^2}{\Delta} <(F^{TF})^2>_\Delta  \int \frac{\Delta
R(\epsilon)}{\epsilon}d\epsilon\right ].   \label{polCE2} \end{equation}

where $<(F^{TF})^2>_\Delta$ is the average of $|F_{\alpha,\beta}^{TF}|^2$ for
$|\epsilon_{\alpha} -\epsilon_\beta | <\Delta$ and $R(\epsilon )$ is the
probability of finding two levels separated by $\epsilon$ whose expression for
the different symmetry classes are known from random matrix theory
\cite{metha}. It is straight forward to show that:

\begin{equation}  \nonumber \int (\frac{ R(\epsilon)^{GUE}}{\epsilon} -
\frac{R(\epsilon)^{GOE}}{\epsilon})d\epsilon= - \frac{1}{2\Delta}
\end{equation}

 The flux dependence of the average matrix element $\delta <(F^{TF})^2>_\Delta
$ is of the order of $\displaystyle -\frac \Delta {E_c}\delta <(F{\alpha ,
\alpha}^{TF})^2>$ as a result:

\begin{equation} \delta \alpha^{CE} \simeq \delta \alpha^{GCE} \frac{\Delta}{E_c}
\label{eq} \end{equation}

  In the limit where $g= E_c/\Delta >> 1$, the level repulsion contribution to
the magneto-polarisability  compensates the flux dependence of $\alpha_D
^{GCE}$ in eq.\ref{polCE2}. As a result the flux dependence of the
polarisability in the CE is strongly reduced compared to the GCE at zero
temperature.  However it is not obvious whether this result obtained with
single electron wave functions is going to survive if electron-electron
interactions are taken into account.  These differences between CE and GCE
averages are also expected to disappear at $T \geq E_c$.

  We have estimated  quantum corrections to the average electrical dynamical
polarisability of 2D Aharonov-Bohm rings which give rise to a low field
positive magneto-polarisability. This effect has been shown to be directly
related to the flux dependence of the matrix elements of the screened potential
in the rings.  In the CEat low temperature the 
magneto-polarisability is sensitive to level spacing statistics whose
contribution cancels the flux dependence due to matrix elements, it is strongly
 reduced  compared to the GCE value. It is possible to evaluate the
magneto-polarisability of a typical GaAs/GaAlAs ring to be of the order of:
$\displaystyle \frac {\delta \alpha}{\alpha}= 3  10^{-3}$. This result is
encouraging for experiments at least when the limit $\omega > \gamma$ is
achieved.

We are grateful to K.B. Efetov who sent us a copy of his work before
publication. We also aknowlege fruitful discussions with M.Bttiker, Y.Imry,
S.Ketteman, G.Montambaux. Numerical simulations have been performed using CRAY facilities at IDRIS (Orsay).

\begin{figure}
	\caption{Flux dependent dynamical $\alpha_D$ and thermodynamical $\alpha_T$
polarisability in the GCE ensemble. These data are obtained from numerical
simulations on the Anderson model for a ring $80 \times 8 $ and  2 different
values of  disorder: $W_d=1$ and $W_d=2$ in the diffusive regime. The vertical
axis has been arbitrarly shifted for the different curves.
   \label{figxphi}}
	\end{figure}

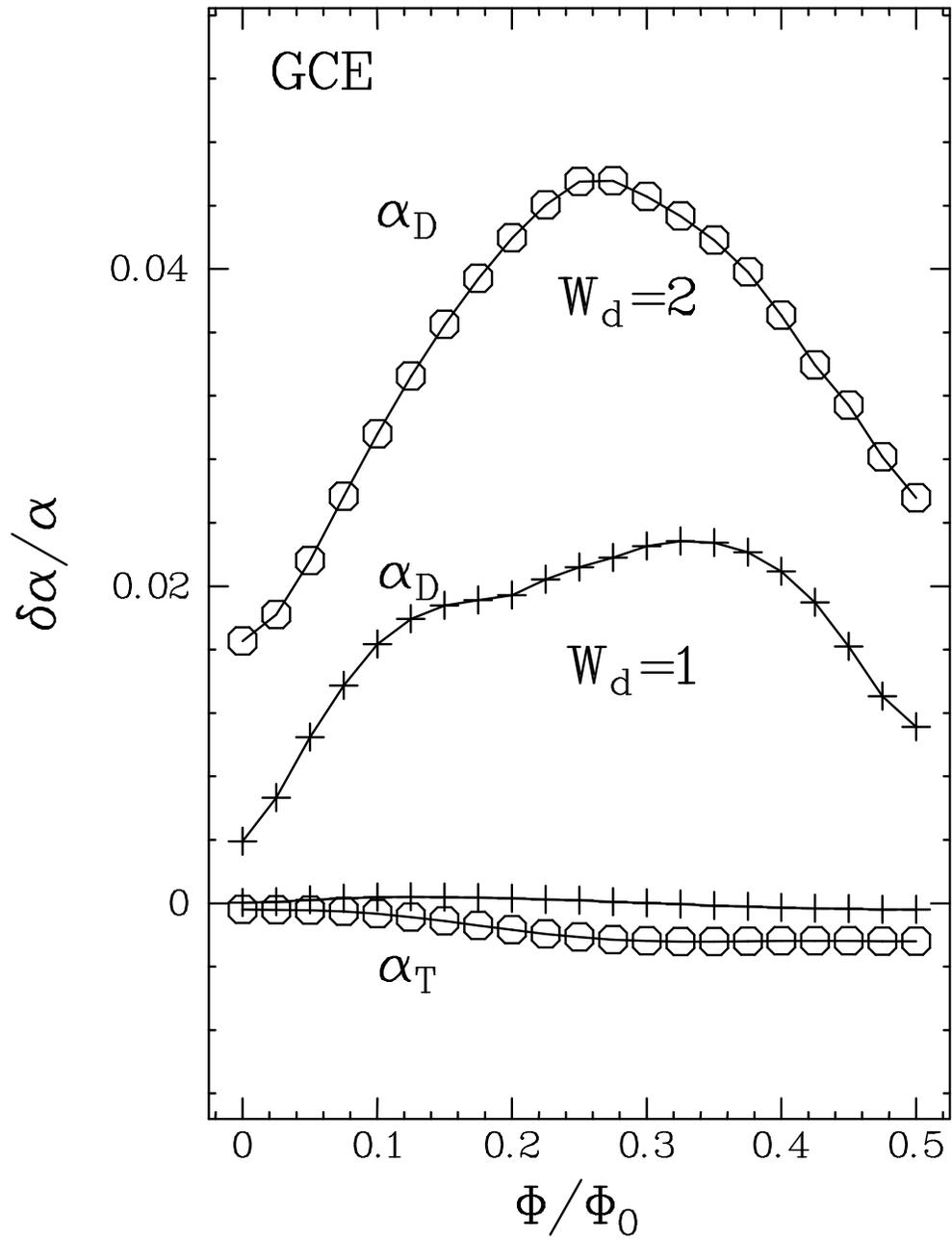
\end{document}